# Experimental Observation of Acoustic Weyl Points and Topological Surface States

Hao Ge[1], Xu Ni[1], Yuan Tian[1], Samit Kumar Gupta[1], Ming-Hui Lu[1, 4*], Xin Lin[2*], Wei-Dong Huang[2], C.T. Chan[3] and Yan-Feng Chen[1, 4]

[1] *Department of Materials Science and Engineering, College of Engineering and Applied Sciences and National Laboratory of Solid State Microstructures, Nanjing University, Nanjing 210093, China*

[2] *State Key Laboratory of Solidification Processing, Northwestern Polytechnic University, Xi'an 710072, China*

[3] *Department of Physics, Hong Kong University of Science and Technology, Clear Water Bay, Hong Kong, China*

[4] *Collaborative Innovation Center of Advanced Microstructures, Nanjing University, Nanjing, Jiangsu 210093, China*

[*] Correspondence should be addressed to: luminghui@nju.edu.cn, xlin@nwpu.edu.cn.

**Abstract**

Weyl points emerge as topological monopoles of Berry flux in the three-dimensional (3D) momentum space and have been extensively studied in topological semimetals. As the underlying topological principles apply to any type of waves under periodic boundary conditions, Weyl points can also be realized in classical wave systems, which are easier to engineer compared to condensed matter materials. Here, we made an acoustic Weyl phononic crystal by breaking space inversion (P) symmetry using a combination of slanted acoustic waveguides. We conducted angle-resolved transmission measurements to characterize the acoustic Weyl points. We also experimentally confirmed the existence of acoustic "Fermi arcs" and demonstrated robust one-way acoustic transport, where the surface waves can overcome a step barrier without reflection. This work lays a solid foundation for the basic research in 3D topological acoustic effects.

**Introduction**

Topological states in acoustic systems are drawing considerable attention and have been studied both in theory and experiment [1-8]. Acoustic topological states that are immune to imperfections offer promising applications, such as one-way wave-guiding, high-Q resonators and sound energy harvesting. However, till now, the research on topological acoustics mainly focuses on two-dimensional (2D) systems. 3D acoustic topological crystals remain to be further explored.

Weyl fermions were originally predicted as the massless solution of the Dirac equation [9], and have recently been discovered in Weyl semimetals [10-14], where Weyl Hamiltonian can be used to describe the low-energy excitations. In Weyl semimetals, Weyl points are point degeneracies at the crossing of two linear dispersion bands in the 3D momentum space, and perturbations of the Weyl Hamiltonian that preserve translational symmetry will not open a gap [15,16]. Weyl points can be regarded as monopoles of Berry flux in momentum space, and their topological charges can be defined by the topological invariants of Chern numbers [10]. A series of unusual effects have been demonstrated in Weyl semimetals, such as chiral anomaly [17], topological surface states [18], and quantum anomalous Hall effect [19]. Band structures carrying Weyl points can also be observed in non-electronic systems [20-28]. Weyl photonic crystals have been realized using parity (P, inversion)-symmetry-broken double-gyroid structures [20,21], and other nodal point structures, such as 3D Dirac points [29], have also been realized in photonics. The research on acoustic Weyl points is also under way, including the theoretical study of type-I and type-II Weyl points [30,31], which can be realized in 3D phononic crystals consisting of waveguides and resonators.

A pioneering experimental work associated with acoustic Weyl semimetal has been reported very recently [32], demonstrating that acoustic Weyl crystals can be realized in practice. Many intriguing phenomena associated with Weyl points and surface states can be explored using acoustic crystals, and in particular, the nodal point characteristics and wavevector selectivity of acoustic Weyl points remain to be further explored. Here, we experimentally observe acoustic Weyl points by performing angle-resolved transmission measurements through P-symmetry-broken 3D phononic crystals. The experimental results clearly exhibit the momentum space nodal point characteristics of Weyl points. We also demonstrate the existence of acoustic "Fermi arcs", which connect pairs of Weyl points. A unique straight-line shape of the "Fermi arcs" was observed, which can result in a collimated and robust propagating surface wave. More importantly, we show that the dispersion of surface states can be tailored by changing the boundary conditions. By constructing a step barrier and scanning the propagating acoustic field inside the air layer, the robust sound transport along the surface has been experimentally confirmed. These results will pave the way for further development of 3D topological acoustics, and provide new strategies for wave manipulation in 3D space.

**Results**

Since the Berry curvature must vanish in the case of PT (time reversal)-symmetry, Weyl points can only exist in the PT-symmetry-broken case, implying that either P or T must be broken. To break the T symmetry of an acoustic system usually requires the introduction of circulating fluids [33] or magneto-acoustic effects [34], which is quite difficult to realize in experiments. We choose to adopt the breaking of P-symmetry to obtain a PT-broken acoustic system. In Fig. 1(a), we show the P-broken phononic crystal having a honeycomb lattice, and the unit cell consisting of a cylinder and a hexagonal prism containing six slanted waveguides. The Brillouin zone is shown in Fig. 1(b). The total number of pairs of Weyl points is two, which is the minimum number that can exist when T-symmetry is preserved. The band structure demonstrates linear dispersion relations around the points K and H [see Fig. 1(c)]. Although the Weyl points at K and H are close in the frequency range, they are separated in the momentum space, making the measurements of both Weyl points feasible.

Two samples are fabricated by 3D printing with the same body structure but different crystal plane to measure the Weyl points at K and H respectively [35]. We conduct angle-resolved transmission measurements of the phononic crystal sample to obtain the 3D dispersions of the bulk states. The illustration of the experimental apparatus is shown in Fig. 2(a). Acoustic plane wave is incident on the surface of the sample, and the angle of incidence can be adjusted by rotating the sample along the vertical axis. As shown in Fig. 2(b), the incident wavevector component parallel to the sample surface can excite all the bulk states with the same frequency and wavevector projection. Since the wavenumber of airborne sound at the Weyl point's frequency is smaller than $\Gamma K$, i.e. the Bloch mode at K will not be excited even in the case of grazing incidence, we choose to excite the Weyl point at K' instead of K. In the scanning process, the angle of $\theta$ and $\alpha$ is adjusted, as illustrated in Fig. 2(a) and 2(b). The four sets of experimental data with different $\alpha$ is shown in Fig. 2(d), where the range of the transmission data is limited due to the maximum rotation angle of $\theta$. In comparison, the calculated band structure is plotted in the inset, where the shaded region corresponds to the frequency range of the experimental measurement. In principle, all the states shown in the projection bands should be observed in the transmission data while the transmission amplitude should be proportional to the density of states. In addition, the coupling and transmission efficiencies depend on several factors such as the details of Bloch modes, mode symmetries, the finite size of the sample, group velocity, etc. Although the signal intensity varies in some regions of the transmission spectrum, the experimental data agrees well with the theoretical predictions. In the case of $\alpha = 0°$, the Weyl point at K' [see Fig. 2(b)] can be excited. The profile of transmission intensity (Fig. 2d1) clearly shows linear dispersions around a point degeneracy. When the angle of $\alpha$ is increased from $0°$ to $15°$, $30°$, and $45°$, we observe the opening of a bandgap around the Weyl point at K'. We note that this bandgap is opened by a point degeneracy instead of a line degeneracy (line node [21]). In our phononic crystal, acoustic wave does not possess spin or polarization degeneracies, and the symmetry group is still symmorphic, therefore excluding the possibility of four-fold 3D Dirac point degeneracies. Similarly, another Weyl point at H [the magenta sphere in Fig. 2(c)] can be studied by rotating the angle $\beta$ [Fig. 2(e)]. Although the transmission intensity is weak in some regions of the upper bulk band, the dispersion of Weyl point can still be inferred.

Since Weyl points are monopoles of quantized Berry flux in momentum space, a 2D plane cut with a fixed $k_z$ ($k_z \neq 0$ and $k_z \neq \pi/(h_p + h_c)$) in 3D momentum space can acquire a net Berry flux, resulting in a non-zero Chern number for the 2D subsystem. The non-zero Chern number implies the existence of topological surface states at the interface between the 3D phononic crystal and a topological trivial system. Here, we consider a 3D phononic crystal that is finite in the $y$ direction and periodic in the $x$ and $z$ direction. The hard boundary condition is applied in the $y$ direction and can be treated as a trivial band gap. We calculate the projection band along the $x$ direction with a fixed $k_z = 0.5\pi/(h_p + h_c)$, and we observe that surface states emerge in the band gap [see Fig. 3(a)]. The red and blue curves correspond to the surface states localized at the lower and upper boundaries, respectively. The calculated eigenpressure fields of the surface states, corresponding to the two black stars in Fig. 3(a), are shown in Fig. 3(b).

The surface Fermi arc dispersion is an important feature of Weyl semimetals. Although there is no "Fermi energy" in acoustic systems, we can trace out the trajectories of surface states at a fixed frequency (near the Weyl point frequency) in the 2D Brillouin zone spanned by $k_x$ and $k_z$. To excite all the surface states on the $x-z$ plane, we inject the sound field into the interface through a 4mm diameter hole at the center of the rigid plate [see Fig. 3(e)]. The rigid plate here serves as the hard boundary. The frequency of the input sound is 13.5 kHz, which is close to the frequency of the Weyl point at H. The simulated acoustic field distribution near the interface [see Fig. 3(c)] agrees well with the experimental result [see Fig. 3(d)], showing that the surface waves propagate along the $x$ axis without broadening, which is the collimation phenomenon of sound. By Fourier transforming the measured acoustic field, we can arrive at the trajectories of surface states, or the acoustic "Fermi arcs", in the 2D Brillouin zone [see Fig. 3(f)]. The white dots denote the calculated acoustic "Fermi arcs". The magenta/green spheres represent the projections of Weyl points at H/K in the 2D Brillouin zone. We see that the acoustic "Fermi arcs" connect the projections of pairs of Weyl points. In phononic crystals, the direction of group velocity is determined by the normal vector to the equi-frequency contour. The "Fermi arcs" here are the equi-frequency contours at the frequency of 13.5 kHz and are nearly straight lines along the $k_z$ direction. The direction of the corresponding group velocity is hence nearly along the $k_x$ direction, which explains the collimation phenomenon at the interface. The surface waves are also topologically protected, therefore we can achieve the collimation phenomenon and robust sound transport at the same time (see simulated result in Fig. S3 [35]).

The dispersion of the surface states is sensitive to the boundary conditions [36], and can be modified by cutting or adding an air layer on the boundary, but the surface states will remain gapless. We calculate the projection bands on the $y-z$ plane with different boundary conditions and the wavevector $k_z$ is fixed as $0.4\pi/(h_p+h_c)$. From Fig. 4(a) and 4(b) we can see that, for the truncated phononic crystal, the surface bands split and two flat branches emerge near the band edge. In Fig. 4(d)-(f), the flat boundary is gradually moved away from the truncated phononic crystal; with the increasing air-layer thickness, the coupling between the waveguide modes and the surface states "compresses" the surface bands. The surface band dispersions are quite tunable, while the net number of one-way modes localized on one side of the sample remains the same. The introduction of air layer on the surface not only provides the feasibility to tailor the surface dispersions, but also makes it easier to generate and probe the surface waves inside the air gap, and therefore making the Weyl phononic crystal prominent practical platform to be applied to potential applications.

In addition, we construct a step barrier to examine the topologically protected one-way sound transport along the surface [schematic see Fig. 5(a)]. The rigid plate is placed at a distance of 10mm from the phononic crystal and the sound is injected through a slanted waveguide placed at the bottom of the air layer. The frequency of the input sound is 13.5 kHz. We scan the field distribution inside the air layer and observe that the surface waves can pass through the step barrier without being reflected at the edge [see Fig. 5(b)], which confirms the topological protection of the surface states. A drop of intensity is noticed, which is due to several factors, such as absorption in air, the finite size of the sample, and beam broadening during propagation. For comparison, we replace the phononic crystal with a rigid plate and scan the field distribution again. The propagating waves are almost entirely specularly reflected at the edge of the step barrier [see Fig. 5(c)].

**Conclusion**

In conclusion, the acoustic Weyl points and "Fermi arcs" observed in this experiment show that we can design, fabricate and characterize topological acoustic crystals whose novel wave transport phenomena are due entirely to the inherent microstructure. This paves the way for the further development of 3D topological acoustics, where materials exhibiting novel features such as 3D Dirac points [37,38] and various gapped topological phases [29,39-41] are yet to be realized in practice. Weyl phononic crystal and its structural designs provide new strategies for sound manipulation in 3D space. Since Weyl point is a single point in the 3D momentum space, with sound incidence of different angles, the transmitted sound will have a specific outgoing direction due to the wavevector selectivity of Weyl point, and can be used in spatial filtering [42]. On the other hand, the topologically protected surface states can be useful for wave-guiding and energy harvesting.

**Acknowledgements**

This work is supported by the National Key R&D Program of China (2017YFA0303702), and the National Natural Science Foundation of China (Grant No. 11625418, No. 51732006, No. 11474158, No. 11604145, and No. 51721001). We also acknowledge the support of China Postdoctoral Science Foundation (Grant No. 2016M591818) and Jiangsu Postdoctoral Science Foundation (Grant No. 1601169C). Work done in Hong Kong is supported by Hong Kong Research Grant Council (Grant AoE/P-02/12).

H. G. and X. N. contributed equally to this work.

**Figure captions**

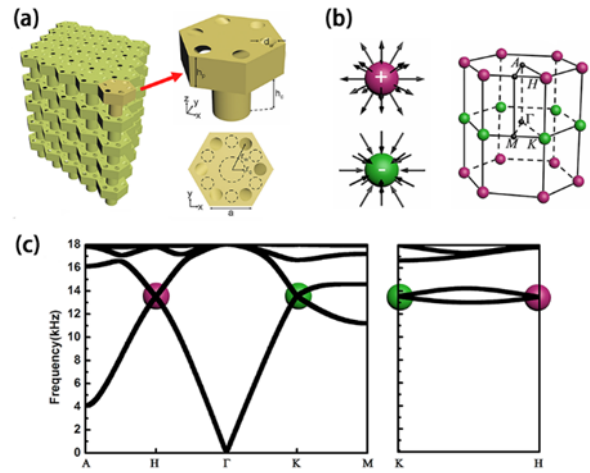

FIG. 1. Weyl points of a P-broken 3D phononic crystal. (a) Illustration of the 3D Weyl phononic crystal with a honeycomb lattice. (b) Schematic of the Brillouin zone of the P-broken 3D phononic crystal, where the magenta/green sphere represents the Weyl point with positive/negative topological charge. (c) The bulk band structure of the phononic crystal, which demonstrates linear dispersion relations around the points K and H.

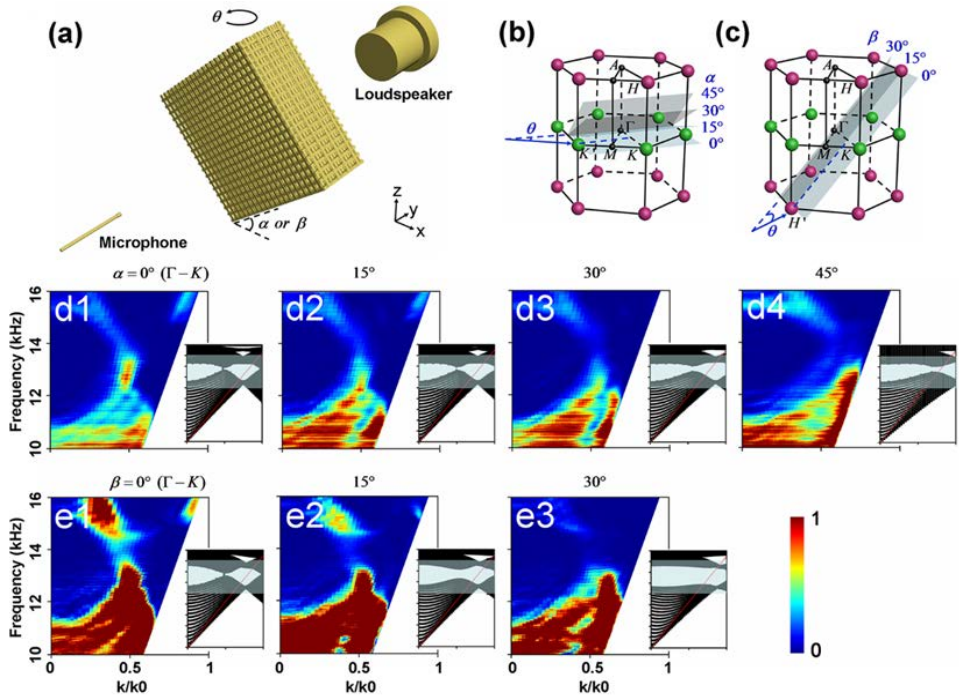

FIG. 2. Experimental observation of acoustic Weyl points. (a) Illustration of the experimental setup. In the measurement, the sample is rotated by the angle $\theta$ along the vertical axis, and the angles $\alpha$ or $\beta$ along a horizontal axis perpendicular to the sample surface. (b)&(c) The illustration of the rotation angle in the Brillouin zone of the 3D phononic crystal sample to measure the Weyl point of (b) K and (c) H. (d)&(e) The measured projection bands of the phononic crystal, where the color from red to blue represents the normalized transmission intensity from 1 to 0. The Weyl points at (d) K and (e) H are measured respectively with different rotation angles. The insets show the calculated band structure.

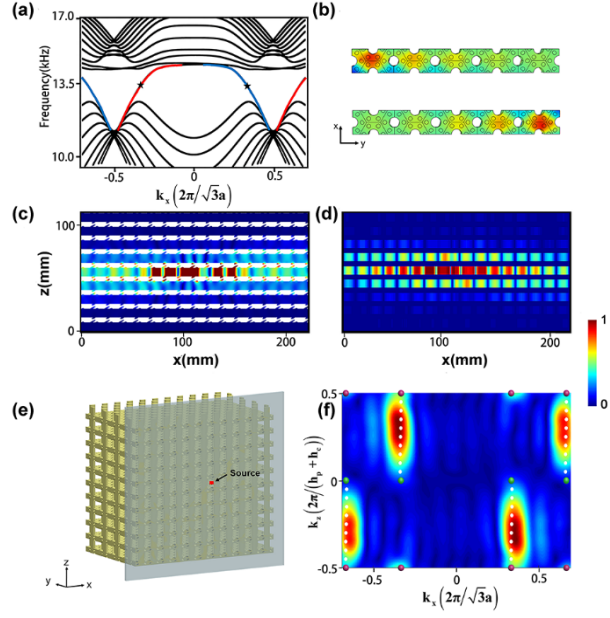

FIG. 3. Surface states and acoustic "Fermi arcs". (a) The calculated projection band along the direction $x$ with a fixed $k_z = 0.5\pi/(h_p + h_c)$. The red and blue curves correspond to the surface states localized at the lower and upper boundaries, respectively. (b) The calculated eigenpressure fields of the surface states at the frequency of 13.5 kHz, corresponding to the two black stars in Fig. 3(a). (c)&(d) The simulated (c) and measured (d) field distribution near the interface. The color from red to blue represents the normalized intensity from 1 to 0. (e) The schematic of the sample. The 3D phononic crystal is covered by a rigid plate and the point source (the red dot) at the center of the interface excites the surface states on the $x-z$ plane. (f) Fourier transform result of the measured field distribution, which corresponds well with the calculated "Fermi arcs" (the white dots). The color from red to blue represents the normalized intensity from 1 to 0.

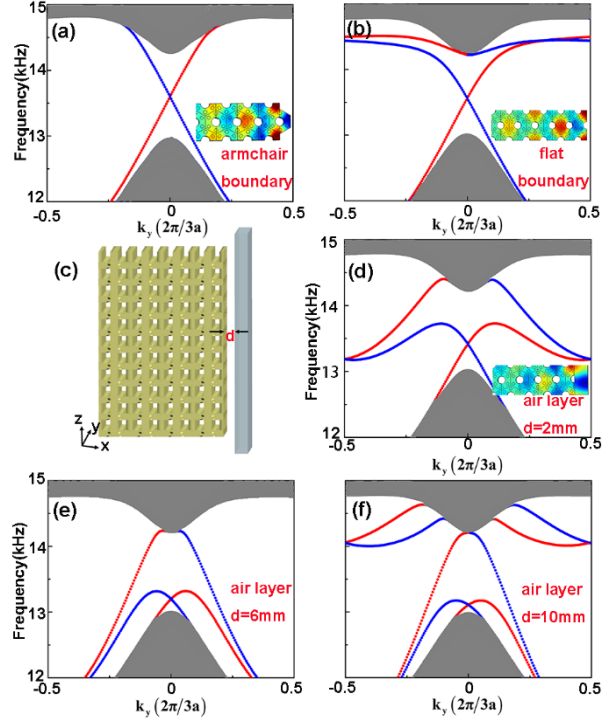

FIG. 4. Influence of boundary conditions. We calculate the projection bands on the $y-z$ plane with a fixed $k_z = 0.4\pi/(h_p + h_c)$. The red and blue curves correspond to the surface states localized at the lower and upper boundaries, respectively. (a)&(b) The projection band with an armchair boundary (a) and a flat boundary (b). (c) The air layer is located between the rigid plate and the truncated phononic crystal. (d)-(f) The projection bands with a 2mm, 6mm, 10mm thickness air layer, respectively.

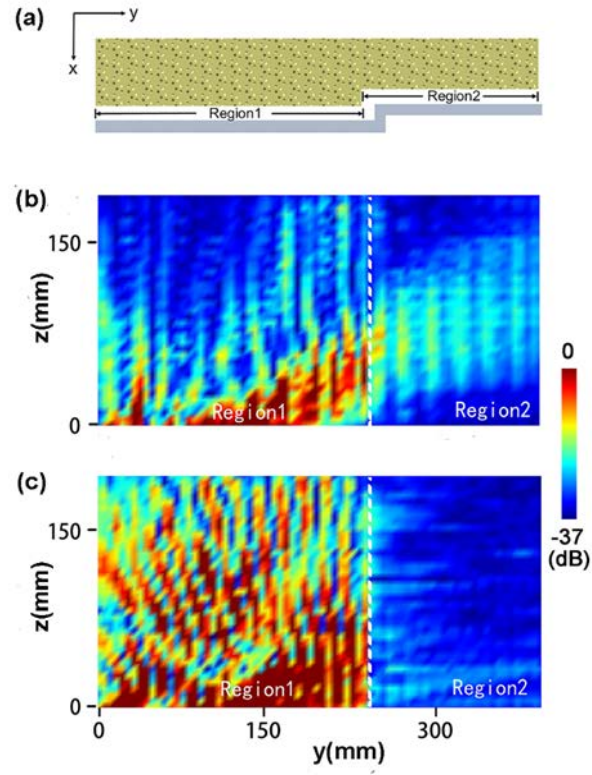

FIG. 5. Topologically protected surface states inside the air layer. (a) Top view of the experimental setup. The sound is injected through a slanted waveguide placed at the bottom of region1. (b) The measured acoustic field distribution on the $y-z$ plane. The white dashed line denotes the edge of the step. The surface waves pass through the step barrier without being reflected. (c) The measured acoustic field distribution inside the step-shape air layer between two rigid plates. Specularly reflection occurs at the edge of the step.